\documentclass[runningheads]{llncs}

\usepackage{amsfonts}
\usepackage{amssymb}
\usepackage{epsfig}
\usepackage{color}

\begin{document}

\title{Triadic motifs and dyadic self-organization in the World Trade Network}
\titlerunning{Triadic Motifs in the World Trade Network}
\author{Tiziano Squartini\inst{1, 2, 3} \and Diego Garlaschelli\inst{1}}
\institute{Instituut-Lorentz for Theoretical Physics, Leiden Institute of Physics, \\University of Leiden, Niels Bohrweg 2, 2333 CA Leiden (The Netherlands)
\and 
Department of Physics, University of Siena, via Roma 56, 53100 Siena (Italy)
\and 
Center for the Study of Complex Systems, via Roma 56, 53100 Siena (Italy)}

\maketitle

\begin{abstract}
In self-organizing networks, topology and dynamics coevolve in a continuous feedback, without exogenous driving. 
The World Trade Network (WTN) is one of the few empirically well documented examples of self-organizing networks: its topology strongly  depends on the GDP of world countries, which in turn depends on the structure of trade.
Therefore, understanding which are the key topological properties of the WTN that deviate from randomness provides direct empirical information about the structural effects of self-organization.
Here, using an analytical pattern-detection method that we have recently proposed, we study the occurrence of triadic `motifs' (subgraphs of three vertices) in the WTN between 1950 and 2000.
We find that, unlike other properties, motifs are not explained by only the in- and out-degree sequences. 
By contrast, they are completely explained if also the numbers of reciprocal edges are taken into account.
This implies that the self-organization process underlying the evolution of the WTN is almost completely encoded into the dyadic structure, which strongly depends on reciprocity.
\end{abstract}

\section{Introduction}
\begin{figure}[t!]
\centering
\includegraphics[width=0.915\textwidth]{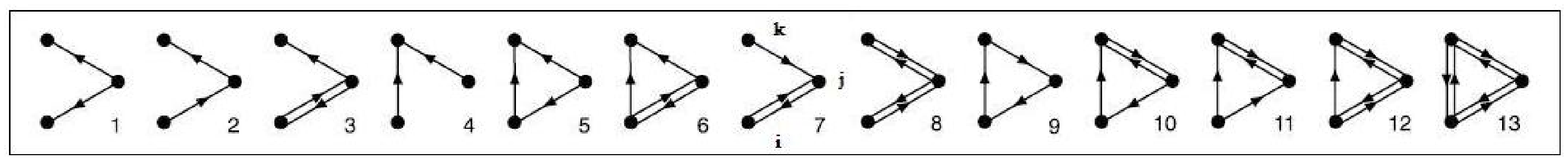}
\caption{The triadic, binary, directed motifs}
\label{mot_img}
\end{figure}

The global economy is a prototypic example of complex self-organizing system, whose collective properties emerge spontaneously through many local interactions. In particular, international trade between countries defines a complex network which arises as the combination of many independent choices of firms. It was shown that the topology of the World Trade Network (WTN) strongly depends on the Gross Domestic Product (GDP) of world countries \cite{garlaWTW}. On the other hand, the GDP depends on international trade by definition \cite{garlaWTW2}, which implies that the WTN is a remarkably well documented example of adaptive network, where dynamics and topology coevolve in a continuous feedback. In general, understanding self-organizing networks is a major challenge for science, as only few models of such networks are analytically solvable \cite{garla_self}. However, in the particular case of the WTN, the binary topology of the network is found to be extremely well reproduced by a null model which incorporates the degree sequence \cite{SFG_WTW1}. These results, which have been obtained using a fast network randomization method that we have recently proposed \cite{SG_method}, make the WTN particularly interesting. 
In this paper, after briefly reviewing our randomization method, we apply it to study the occurrence of triadic `motifs', i.e. directed patterns involving three vertices (see Fig.\ref{mot_img}).
We show that, unlike other properties which have been studied elsewhere \cite{SFG_WTW1}, the occurrence of motifs is not explained by only the in- and out-degrees of vertices. 
However, if also the numbers of reciprocal links of each vertex (the \emph{reciprocal degree sequence}) are taken into account, the occurrences of triadic motifs are almost completely reproduced. 
This implies that, if local information is enhanced in order to take into account the reciprocity structure, motifs display no significant deviations from random expectations.
Therefore the (in principle complicated) self-organization process underlying the evolution of the WTN turns out to be relatively simply encoded into the local dyadic structure, which separately specifies the number of reciprocated and non-reciprocated links per vertex.
Thus the dyadic structure appears to carry a large amount of information about the system.

\section{Searching for Non-Random Patterns in the WTN}
In this section we briefly summarize our recently proposed randomization method and how it can be used to detect patterns when local constraints are considered. 

\subsection{Network Pattern Detection: the Randomization Method\label{sec_method}}
Our method, which is based on the maximum-likelihood estimation of maximum-entropy models of graphs, introduces a family of null models of a real network and uses it to detect topological patterns analytically \cite{SG_method}. 
Defining a null model means setting up a method to assign probabilites \cite{MSZ,HL,newman_expo}. 
In our approach, a real network $\mathbf{G}^{*}$ with $N$ vertices is given (either a binary or a weighted graph, and either directed or undirected, whose generic entry is $g_{ij}$) and a way to generate a family $\mathcal{G}$ of randomized variants of $\mathbf{G}^{*}$ is provided, by assigning each graph $\mathbf{G}\in\mathcal{G}$ a probability $P(\mathbf{G})$. 
In the method, the probabilities $P(\mathbf{G})$ are such that a maximally random ensemble of networks is generated, under the constraint that, on average, a set $\{C_{a}\}$ of desired topological properties is set equal to the values $\{C_{a}(\mathbf{G}^*)\}$ observed in the real network $\mathbf{G}^*$.
This is achieved as the result of a constrained Shannon-Gibbs entropy maximization \cite{newman_expo}

\begin{equation}
S\equiv -\sum_\mathbf{G} P(\mathbf{G})\ln P(\mathbf{G})
\label{eq_entropy}
\end{equation}

\noindent and the imposed constraints are the normalization of the probability and the average values of a number of chosen properties, $\{C_{a}\}$:

\begin{equation}
1=\sum_{\mathbf{G}}P(\mathbf{G}),\,\langle C_{a}\rangle\equiv\sum_\mathbf{G} C_{a}(\mathbf{G}) P(\mathbf{G})\enspace .
\end{equation}

\noindent This optimization leads to exponential probability coefficients \cite{newman_expo}

\begin{equation}
P(\mathbf{G})=\frac{e^{-H(\mathbf{G},\vec{\theta})}}{Z}
\end{equation}

\noindent where the linear combination of the contraints $H(\mathbf{G},\vec{\theta})\equiv\sum_a \theta_a C_a(\mathbf{G})$ is called \emph{graph hamiltonian} (the coefficients $\{\theta_a\}$ are free parameters, acting as Lagrange multipliers controlling the expected values $\{\langle C_a\rangle\}$) and the denominator $Z\equiv\sum_\mathbf{G} e^{-H(\mathbf{G},\vec{\theta})}$ is called \emph{partition function}. 

The next step is the maximization of the probability $P(\mathbf{G^*})$ to obtain the observed graph $\mathbf{G^*}$, i.e. the real-world network to randomize \cite{SG_method}. This step fixes the values of the Lagrange multipliers as they are found by maximizing the log-likelihood

\begin{equation}
\mathcal{L}(\vec{\theta})\equiv \ln P(\mathbf{G^*}|\vec{\theta})=-H(\mathbf{G^*},\vec{\theta})-\ln Z(\vec{\theta})
\label{eq_likelihood}
\end{equation}

\noindent to obtain the real network $\mathbf{G^*}$. It can be easily verified \cite{mylikelihood} that this is achieved by the parameter values $\vec{\theta}^*$ satisfying

\begin{equation}
\langle C_a\rangle^*=\sum_{\mathbf{G}}C_{a}(\mathbf{G})P(\mathbf{G}|\vec{\theta}^*)=C_a(\mathbf{G^*})\quad \forall a
\label{eq_caverage}
\end{equation}

\noindent that is, that the ensemble average of each constraint, $\langle C_{a}\rangle$, equals the observed value on the real network, $C_{a}(\mathbf{G}^*)$. Once the numerical values of the Lagrange multipliers are found, they can be used to find the ensemble average $\langle X\rangle^*$ of any topological property $X$ of interest \cite{SG_method}:

\begin{equation}
\langle X\rangle^*=\sum_\mathbf{G}X(\mathbf{G})P(\mathbf{G}|\vec{\theta}^*)\enspace .
\label{eq_X*}
\end{equation}

\noindent The exact computation of the expected values can be very difficult. For this reason it is often necessary to rest on the \emph{linear approximation method} \cite{SG_method}. However, in the present study we will consider particular topological properties $X$ (i.e. motif counts, see below) whose expected value can be evaluated \emph{exactly}. Our method also allows to obtain the variance of $X$ by applying the usual definition:

\begin{equation}
\sigma^2[X]=\langle[X(\mathbf{G})-\langle X\rangle)]^2\rangle=\sum_{i,j}\sum_{t,s}\sigma[g_{ij},g_{ts}]\left(\frac{\partial X}{\partial g_{ij}}\frac{\partial X}{\partial g_{ts}}\right)_{\mathbf{G}=\langle\mathbf{G}\rangle}
\label{eq_generalpropagation}
\end{equation}

\noindent where $\sigma[g_{ij},g_{ts}]$ is the covariance of the adjacency matrix elements $g_{ij}$ and $g_{ts}$. This formula can be greatly simplified by considering probabilities that are factorizable in terms of dyadic probabilities, as follows

\begin{equation}
P(\mathbf{G}|\vec{\theta})=\prod_{i<j}D_{ij}(g_{ij},g_{ji}|\vec{\theta})
\label{eq_factorize}
\end{equation}

\noindent where the product runs over all the dyads, that is the unordered pairs of vertices $(i,j)$ (with $i<j$), and $D_{ij}(g,g'|\vec{\theta})$ is the joint probability that $g_{ij}=g$ and $g_{ji}=g'$. Finally, the variance of $X$ evaluated in $\vec{\theta}^*$ becomes

\begin{equation}
(\sigma^*[X])^2=\sum_{i,j}\left[\left(\sigma^*[g_{ij}]\frac{\partial X}{\partial g_{ij}}\right)_{\mathbf{G}=\langle\mathbf{G}\rangle^*}^2+\sigma^*[g_{ij},g_{ji}]\left(\frac{\partial X}{\partial g_{ij}}\frac{\partial X}{\partial g_{ji}}\right)_{\mathbf{G}=\langle\mathbf{G}\rangle^*}\right]\enspace .
\end{equation}

The joint knowledge of $\langle X\rangle^*$ and $\sigma^*[X]$ allows to detect deviations from randomness in the observed topology. In particular, as we show later, it is possible to calculate by how many standard deviations the observed value $X^*$ differs from the expected value $\langle X\rangle^*$. 
Quantities which are consistent with their expected value are explained by the enforced constraints $\{C_{a}\}$. On the other hand, significantly deviating properies cannot be traced back to the constraints and therefore signal the incompleteness of the information encoded in the constraints.
Other approaches achieve this result by explicitly generating many randomized variants of the real network,  measuring $X$ on each such variant, and finally computing the sample average and standard deviation of $X$ \cite{MSZ}. This is extremely time consuming, especially for complicated topological properties.
By contrast, our method is entirely analytical. It yields any expected quantity $\langle X\rangle^*$ in a time as short as that required in order to measure $X^*$ on the single network $\mathbf{G}^*$ \cite{SG_method}.

\subsection{The Role of Local Constraints}
If the network is a binary graph (i.e. if each graph $\mathbf{G}$ in the ensemble is uniquely specified by its adjacency matrix $\mathbf{A}$), then the simplest (i.e. local) choice of the constraints is the \emph{degree sequence}, i.e. the vector of degrees (numbers of incident links) of all vertices. 
For directed networks, which are our interest here, there are actually two degree sequences: the observed in-degree sequence $\mathbf{k}^{\rm in}(\mathbf{A}^*)$ (with $k_{i}^{\rm in}=\sum_{j\neq i}a_{ji}$) and the observed out-degree sequence $\mathbf{k}^{\rm out}(\mathbf{A}^*)$ (with $k_{i}^{\rm out}=\sum_{j\neq i}a_{ij}$). This null model, which is known as the \emph{directed configuration model} (DCM), can be completely dealt with analytically using our method (see Appendix). 
When applied to the WTN, the DCM shows that many topological properties (such as the degree-degree correlations and the directed clustering coefficients) are in complete accordance with the expectations \cite{SFG_WTW1}. This shows that the degree sequences $\mathbf{k}^{\rm in}$ and $\mathbf{k}^{\rm out}$ are extremely informative, as their (partial) knowledge allows to reconstruct many aspects of the (complete) topology. 
On the other hand, it was also shown that the \emph{reciprocity} of the WTN is highly non-trivial \cite{myreciprocity}. This means that the occurrence of reciprocal links is much higher than expected under any model which, as the DCM, treats two reciprocal links (e.g. $i\to j$ and $j\to i$) as statistically independent. 
A direct consequence is that the reciprocity, as well as any higher-order directional pattern, should not be reproduced by the DCM.
These seemingly conflicting results can only be reconciled if, for some reason, the topological properties that have been studied under the DCM \cite{SFG_WTW1} mask the effects of reciprocity. 
In particular, the directed clustering coefficients, which are based on ratios of realized triangles over the maximum number for each vertex, may show no overall deviation from the DCM, even if the numerator and denominator separately deviate from it. 
In what follows, we investigate this possibility by considering all the observed subgraphs of three vertices (which include both open and closed triangles) separately.
Also, we will use an additional null model which also takes the number of reciprocal links of each vertex into account. 
This second null model is the \emph{reciprocal configuration model} (RCM) \cite{SG_method,myreciprocity,mygrandcanonical}. The local constraints defining it are the three, observed directed-degree sequences $\textbf{k}^{\rightarrow}(\mathbf{A}^*)$, with $k^{\rightarrow}_i\equiv \sum_{j\ne i} a^\rightarrow_{ij}$ (non-reciprocated out degree), $\textbf{k}^{\leftarrow}(\mathbf{A}^*)$, with $k^{\leftarrow}_i\equiv \sum_{j\ne i} a^\leftarrow_{ij}$ (non-reciprocated in-degree) and $\textbf{k}^{\leftrightarrow}(\mathbf{A}^*)\}$, with $k^{\leftrightarrow}_i\equiv \sum_{j\ne i} a^\leftrightarrow_{ij}$ (reciprocated degree) to be imposed across the ensemble of networks having the same number of vertices of the observed configuration and, on average, the above-mentioned directed-degree sequences.
In the Appendix we describe both the DCM and the RCM in more detail, and derive their expectations explicitly.

\subsection{Triadic Motifs in the WTN}
In the following analyses, we  use  yearly  bilateral  data  on  exports  and  imports from the Gleditsch Database\footnote{Gleditsch, K.S.: Expanded trade and GDP data. Jour. Confl. Res. {\bfseries 46} (2002) 712--724.} to analyse the six years 1950, 1960, 1970, 1980, 1990, 2000. This database contains aggregated trade data between countries, i.e. data as they result by summing the single commodity-specific trade exchanges. So we end up with six different, real, asymmetric matrices with entries $m_{ij}^{\rm agg}(y)$ ($y=1950,\,1960\dots 2000$).
These adjacency matrix elements are the fundamental data allowing us to obtain all the possible representations of the WTN: to build the binary, directed representation we are interested in here, we restrict ourselves to consider two different vertices as linked, whenever the corresponding element $m_{ij}^{\rm agg}(y)$ is strictly positive. This implies that the adjacency matrix of the binary, directed representation of the WTN in year $y$ is simply obtained by applying the Heaviside step function to the database entries, i.e. $
a_{ij}(y)=\Theta[m_{ij}^{\rm agg}(y)]$.

Triadic motifs, i.e. all the possibile directed patterns connecting three vertices, are the natural generalizations of directed clustering coefficients \cite{fagiolo_clustering} and the starting point for the understanding of a complex network's self-organization in communities. Thirteen, non-isomorphic, triadic directed patterns can be indentified and classified \cite{calda_book}.
Given a real, binary, directed matrix $\mathbf{A}^*$, the motifs occurrences $N_{m}$ can be written in at least two different ways (see Table \ref{motable}). The first one prescribes to define them in terms of the adjacency matrix entries, $\{a_{ij}\}$. The second one allows to compactly express the coefficients $N_{m}$ by introducing the following dyadic variables

\begin{equation}
a_{ij}^\rightarrow\equiv a_{ij}(1-a_{ji}),\,a_{ij}^\leftarrow\equiv a_{ji}(1-a_{ij}),\,a_{ij}^\leftrightarrow\equiv a_{ij}a_{ji},\,a_{ij}^\nleftrightarrow\equiv (1-a_{ij})(1-a_{ji})
\end{equation}

\noindent thus making the role of reciprocity explicit. However, the number of occurrences of the particular motif $m$ (where $m$ ranges from 1 to 13) as measured on the observed network $\mathbf{A}^{*}$ is uninformative unless a comparison with a properly defined null model is made (see Appendix). This implies that the occurrence of a motif should be compared with its expected value $\langle N_{m}\rangle^{*}$, as computed under the chosen null model. 
This can be compactly achieved by introducing the so-called $z$-score

\begin{equation}
z[N_{m}]\equiv\frac{N_{m}(\mathbf{A}^*)-\langle N_{m}\rangle^*}{\sigma^*[N_{m}]}
\end{equation}

\noindent measuring by how many standard deviations, $\sigma^*[N_{m}]$, the observed and the expected occurrences of motif $m$ differ. Large absolute values of $z[N_{m}]$ indicate motifs that are either over- or under-represented under the particular null model considered and therefore not explained by the constraints defining it, as shown in Fig.\ref{motCM} and Fig.\ref{motRCM} and discussed in the next section.

\begin{table}[t!]
\centering
\caption{Classification (after \cite{calda_book}) and definitions of the triadic motifs}
\begin{tabular}{llllll}
\hline\noalign{\smallskip}
Motif $m$ & $N_{m}$: $1^{st}$ definition & $N_{m}$: $2^{nd}$ definition\\
\noalign{\smallskip}
\hline
\noalign{\smallskip}
$1$ & $\sum_{i\neq j\neq k}(1-a_{ij})a_{ji}a_{jk}(1-a_{kj})(1-a_{ik})(1-a_{ki})$ & $\sum_{i\neq j\neq k}a_{ij}^{\leftarrow}a_{jk}^{\rightarrow}a_{ik}^{\nleftrightarrow}$\\
\hline
$2$ & $\sum_{i\neq j\neq k}a_{ij}(1-a_{ji})a_{jk}(1-a_{kj})(1-a_{ik})(1-a_{ki})$ & $\sum_{i\neq j\neq k}a_{ij}^{\rightarrow}a_{jk}^{\rightarrow}a_{ik}^{\nleftrightarrow}$\\
\hline
$3$ & $\sum_{i\neq j\neq k}a_{ij}a_{ji}a_{jk}(1-a_{kj})(1-a_{ik})(1-a_{ki})$ & $\sum_{i\neq j\neq k}a_{ij}^{\leftrightarrow}a_{jk}^{\rightarrow}a_{ik}^{\nleftrightarrow}$\\
\hline
$4$ & $\sum_{i\neq j\neq k}(1-a_{ij})(1-a_{ji})a_{jk}(1-a_{kj})a_{ik}(1-a_{ki})$ & $\sum_{i\neq j\neq k}a_{ij}^{\nleftrightarrow}a_{jk}^{\rightarrow}a_{ik}^{\rightarrow}$\\
\hline
$5$ & $\sum_{i\neq j\neq k}(1-a_{ij})a_{ji}a_{jk}(1-a_{kj})a_{ik}(1-a_{ki})$ & $\sum_{i\neq j\neq k}a_{ij}^{\leftarrow}a_{jk}^{\rightarrow}a_{ik}^{\rightarrow}$\\
\hline
$6$ & $\sum_{i\neq j\neq k}a_{ij}a_{ji}a_{jk}(1-a_{kj})a_{ik}(1-a_{ki})$ & $\sum_{i\neq j\neq k}a_{ij}^{\leftrightarrow}a_{jk}^{\rightarrow}a_{ik}^{\rightarrow}$\\
\hline
$7$ & $\sum_{i\neq j\neq k}a_{ij}a_{ji}(1-a_{jk})a_{kj}(1-a_{ik})(1-a_{ki})$ & $\sum_{i\neq j\neq k}a_{ij}^{\leftrightarrow}a_{jk}^{\leftarrow}a_{ik}^{\nleftrightarrow}$\\
\hline
$8$ & $\sum_{i\neq j\neq k}a_{ij}a_{ji}a_{jk}a_{kj}(1-a_{ik})(1-a_{ki})$ & $\sum_{i\neq j\neq k}a_{ij}^{\leftrightarrow}a_{jk}^{\leftrightarrow}a_{ik}^{\nleftrightarrow}$\\
\hline
$9$ & $\sum_{i\neq j\neq k}(1-a_{ij})a_{ji}(1-a_{jk})a_{kj}a_{ik}(1-a_{ki})$ & $\sum_{i\neq j\neq k}a_{ij}^{\leftarrow}a_{jk}^{\leftarrow}a_{ik}^{\rightarrow}$\\
\hline
$10$ & $\sum_{i\neq j\neq k}(1-a_{ij})a_{ji}a_{jk}a_{kj}a_{ik}(1-a_{ki})$ & $\sum_{i\neq j\neq k}a_{ij}^{\leftarrow}a_{jk}^{\leftrightarrow}a_{ik}^{\rightarrow}$\\
\hline
$11$ & $\sum_{i\neq j\neq k}a_{ij}(1-a_{ji})a_{jk}a_{kj}a_{ik}(1-a_{ki})$ & $\sum_{i\neq j\neq k}a_{ij}^{\rightarrow}a_{jk}^{\leftrightarrow}a_{ik}^{\rightarrow}$\\
\hline
$12$ & $\sum_{i\neq j\neq k}a_{ij}a_{ji}a_{jk}a_{kj}a_{ik}(1-a_{ki})$ & $\sum_{i\neq j\neq k}a_{ij}^{\leftrightarrow}a_{jk}^{\leftrightarrow}a_{ik}^{\rightarrow}$\\
\hline
$13$ & $\sum_{i\neq j\neq k}a_{ij}a_{ji}a_{jk}a_{kj}a_{ik}a_{ki}$ & $\sum_{i\neq j\neq k}a_{ij}^{\leftrightarrow}a_{jk}^{\leftrightarrow}a_{ik}^{\leftrightarrow}$\\
\hline
\label{motable}
\end{tabular}
\end{table}

\section{Results and Discussion}
\begin{figure}[h!]
\centering
\includegraphics[width=0.8\textwidth]{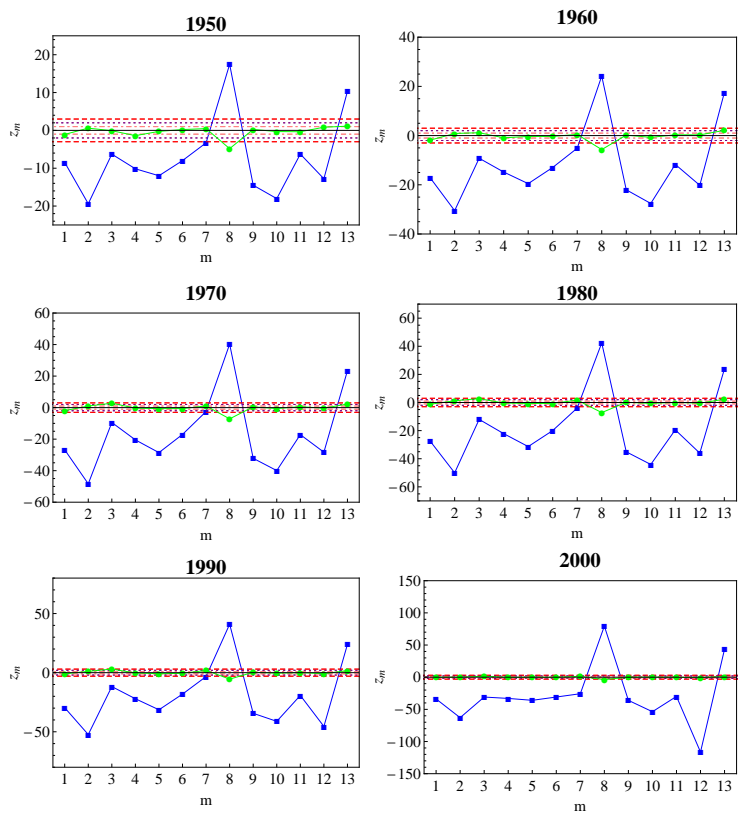}
\caption{$z$-scores of the 13 triadic, binary, directed motifs for the six decades of the WTN, under the DCM ($\textcolor{blue}{\blacksquare}$) and under the RCM ($\textcolor{green}{\bullet}$). The dashed, red lines represent the values $z=\pm3$, the dotted, purple lines the values $z=\pm2$ and the dot-dashed, pink lines the values $z=\pm1$.}
\label{motCM}
\end{figure}

\begin{figure}[h!]
\centering
\includegraphics[width=0.8\textwidth]{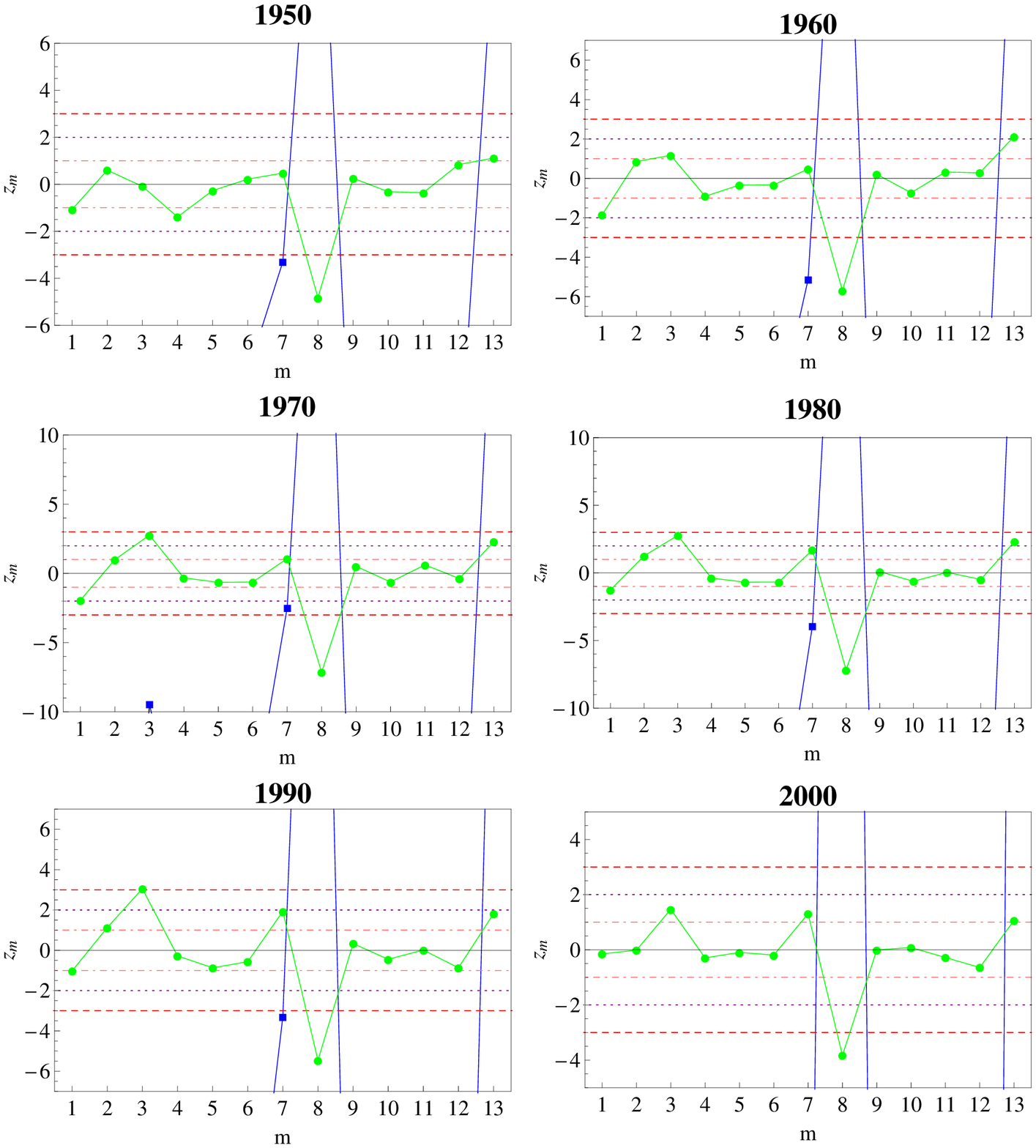}
\caption{$z$-scores of the 13 triadic, binary, directed motifs for the six decades of the WTN, under the DCM ($\textcolor{blue}{\blacksquare}$) and under the RCM ($\textcolor{green}{\bullet}$): zoom of Fig. \ref{motCM}. The dashed, red lines represent the values $z=\pm3$, the dotted, purple lines the values $z=\pm2$ and the dot-dashed, pink lines the values $z=\pm1$.}
\label{motRCM}
\end{figure}

Fig.\ref{motCM} and Fig.\ref{motRCM} show the $z$-scores for all the 13, triadic, binary, directed motifs, computed for the six different snapshots of the World Trade Network corresponding to the decades 1950, 1960, 1970, 1980, 1990, 2000, for both the DCM and the RCM. We also show the six lines $z=\pm1$, $z=\pm2$ and $z=\pm3$ to highlight the region within 3 sigmas from the expectation value.
The analysis reveals a dramatic difference between the predictions of the two null models.  The presence of intrinsically directed trading relationships implies that reciprocity is a fundamental quantity, shaping the network of exchanges among world-countries. 


The reciprocity of the WTN is known to be very high \cite{myreciprocity} and this has strong effects on its motif structure. This implies that, in order to reproduce the topology of the network, it is essential to reproduce its dyadic structure, which in turn strongly depends on reciprocity. This also implies that the RCM should fluctuate less than the DCM, because of this huge amount of additional information. Indeed, this is confirmed by our analysis. 

In fact, we find that, unlike other topological properties \cite{SFG_WTW1}, triadic motifs are systematically not reproduced if only the in-degree and out-degree sequences are taken into account. In particular, the observed motifs counts deviate by up to 100 standard deviations from the expected ones. By contrast, if also the reciprocity (which separately specifies the number of reciprocated links) is introduced in the null model, triadic motifs are almost always consistent with expectations within one standard deviation, as for $m=5,10,11,12$ or at most two standard deviations, as $m=1,4$. Moreover, the significance profiles are almost completely inverted: some of the motifs being over(under)-represented without the reciprocity constraint, as $m=8$ ($m=2$), become under(over)-represented with the reciprocity constraint.

\section{Conclusions}
The WTN is a particularly interesting network which is known to be driven by a self-organization process involving the global economy. 
Empirically, it is a very well documented network which allows to test predictions of null models about the key ingredients shaping its topology.
In this paper, we aimed at isolating the key properties of the WTN topology where the effects of self-organizations can be clearly detected as deviations from randomness.
While the DCM, which takes into account the number of incoming and outgoing connections of all vertices, reproduces many other topological properties, it cannot replicate the observed triadic motifs.
On the other hand, we found that the RCM, which also takes into account the numbers of reciprocated links, can replicate almost perfectly the triadic structure.
We therefore found that the process underlying the evolution of the WTN is mainly encoded into the dyadic structure, which carries a large amount of information about the system: so, the upgrade to the RCM is necessary and the possibility to treat the RCM analytically using our method is therefore an important step forward. The result that dyadic properties almost completely explain triadic ones suggests that in the WTN also higher-order properties (e.g. subgraphs of four or more vertices, and in general even the existence of denser communities of vertices) will be explained mostly by dyadic properties. This raises the important question whether the available community detection methods are successful in identifying communities which are not explained by local constraints. In particular, our results suggest that modularity-based community detection methods will detect communities in the WTN if a null model without reciprocity (DCM) is used, while they should find weak or no community structure if a null model including reciprocity (RCM) is used. In both cases, as we have already claimed \cite{SG_method}, the available expressions for the modularity (which make the strong assumption of  sparse networks) should be revised using the correct expectation values provided by our method, since the WTN is a very dense network. We will address these issues in future work.

\section*{Acknowledgements}
This work was supported by the Dutch Econophysics Foundation (Stichting Econophysics, Leiden, The Netherlands) with funds from beneficiaries of Duyfken Trading Knowledge BV, Amsterdam, The Netherlands.

\section{Appendix}

\subsection{The Directed Configuration Model\label{app_bdn}}

Given a real, binary, directed network $\mathbf{A}^*$ with out-degree and in-degree sequences $\mathbf{k}^{\rm out}(\mathbf{A}^*)$ and $\mathbf{k}^{\rm in}(\mathbf{A}^*)$, the method described in section \ref{sec_method} can be specified in the following way.

\paragraph{The DCM Hamiltonian.} The Hamiltonian implementing the DCM is

\begin{equation}
H(\mathbf{A},\vec{\alpha},\vec{\beta})=\sum_i [\alpha_i k_i^{out}(\mathbf{A})+\beta_i k_i^{in}(\mathbf{A})]=\sum_{i\ne j}(\alpha_i+\beta_j)a_{ij}\enspace;
\end{equation}

\noindent the partition function can be calculated as in \cite{newman_expo}, $Z(\vec{\alpha},\vec{\beta})=\sum_\mathbf{A}e^{-H(\mathbf{A},\vec{\alpha},\vec{\beta})}=\prod_{i\ne j}(1+e^{-\alpha_i-\beta_j})$, and the graph probability is

\begin{equation}
P(\mathbf{A}|\vec{\alpha},\vec{\beta})
\equiv\prod_{i<j}D_{ij}(a_{ij},a_{ji}|\alpha_{i},\alpha_{j},\beta_{i},\beta_{j})=\prod_{i\neq j}p_{ij}^{a_{ij}}(1-p_{ij})^{(1-a_{ij})}
\end{equation}

\noindent and, by setting $x_i\equiv e^{-\alpha_i}$ and $y_i\equiv e^{-\beta_i}$,

\begin{equation}
p_{ij}=\frac{e^{-\alpha_i-\beta_j}}{1+e^{-\alpha_i-\beta_j}}\equiv\frac{x_iy_j}{1+x_iy_j}\enspace .
\label{eq_prob}
\end{equation}

\paragraph{The Log-Likelihood Equations.} The log-likelihood function to maximize is 

\begin{equation}
\mathcal{L}(\mathbf{x},\mathbf{y})=\sum_i\left[ k^{\rm out}_i(\mathbf{A}^*)\ln x_i+k^{\rm in}_i(\mathbf{A}^*)\ln y_i\right]-\sum_{i\ne j}\ln (1+x_i y_j)
\label{eq_likelihoodbdn}
\end{equation}

\noindent and the values $\mathbf{x}^*$, $\mathbf{y}^*$ corresponding to the point of maximum can be found by solving the following system

\begin{eqnarray}
\left\{ \begin{array}{ll}
k^{\rm in}_i(\mathbf{A}^*) &= \sum_{j\ne i}\frac{x^*_iy^*_j}{1+x^*_iy^*_j}\quad \forall i\\
k^{\rm out}_i(\mathbf{A}^*) &= \sum_{j\ne i}\frac{x^*_jy^*_i}{1+x^*_jy^*_i}\quad \forall i
\end{array} \right.
\end{eqnarray}

\paragraph{Expectation Values and Variances.} The expectation value and variance of $a_{ij}$ are, respectively, $\langle a_{ij}\rangle^{*}=p_{ij}^{*}$ and $(\sigma^*[a_{ij}])^2=\langle a_{ij}\rangle^{*}(1-\langle a_{ij}\rangle^*)=p_{ij}^{*}(1-p_{ij}^{*})$. Distinct pairs of vertices are independent random variables; since the first definition of the motifs occurrences only involves products of the adjacency matrix entries, their expectation values can be easily computed, as shown in Table \ref{motable2}. The variance of $N_m$ becomes

\begin{equation}
(\sigma^*[N_{m}])^2=\sum_{i,j}\left[\left(\sigma^*[a_{ij}]\frac{\partial N_{m}}{\partial a_{ij}}\right)_{\mathbf{A}=\langle\mathbf{A}\rangle^*}^2\right]
\end{equation}

\noindent considering that $\sigma^{*}[a_{ij},a_{ji}]=\langle a_{ij}a_{ji}\rangle^{*}-\langle a_{ij}\rangle^{*}\langle a_{ji}\rangle^{*}=0$.

\subsection{The Reciprocal Configuration Model\label{app_rdn}}

Given a real binary directed network $\mathbf{A}^*$ with directed-degree sequences $\mathbf{k}^\rightarrow(\mathbf{A}^*)$, $\mathbf{k}^\leftarrow(\mathbf{A}^*)$ and $\mathbf{k}^\leftrightarrow(\mathbf{A}^*)$ where

\begin{equation}
k_i^\rightarrow(\mathbf{A}^*)\equiv\sum_{j\ne i}a^*_{ij}(1-a^*_{ji}),\,k_i^\leftarrow(\mathbf{A}^*)\equiv\sum_{j\ne i}a^*_{ji}(1-a^*_{ij}),\,k_i^\leftrightarrow(\mathbf{A}^*)\equiv\sum_{j\ne i}a^*_{ij}a^*_{ji}
\end{equation}

\noindent the randomization procedure can be specified in the following way.

\paragraph{The RCM Hamiltonian.} The Hamiltonian implementing the RCM is

\begin{equation}
H(\mathbf{A},\vec{\alpha},\vec{\beta},\vec{\gamma})=\sum_i [\alpha_i k_i^\rightarrow(\mathbf{A})+\beta_i k_i^\leftarrow(\mathbf{A})+\gamma_i k_i^\leftrightarrow(\mathbf{A})]\enspace ,
\nonumber
\end{equation}

\noindent the partition function can be calculated as in \cite{mygrandcanonical}, $Z(\vec{\alpha},\vec{\beta},\vec{\gamma})=\prod_{i< j}(1+e^{-\alpha_i-\beta_j}+e^{-\alpha_j-\beta_i}+e^{-\gamma_i-\gamma_j})$, and the graph probability is

\begin{eqnarray}
P(\mathbf{A}|\vec{\alpha},\vec{\beta},\vec{\gamma})&=&\prod_{i<j}D_{ij}(a_{ij},a_{ji}|
\alpha_{i},\alpha_{j},\beta_{i},\beta_{j},\gamma_{i},\gamma_{j})\nonumber\\
&=&\prod_{i<j}(p^\rightarrow_{ij})^{a^\rightarrow_{ij}}(p^\leftarrow_{ij})^{a^\leftarrow_{ij}}
(p^\leftrightarrow_{ij})^{a^\leftrightarrow_{ij}}
(p^\nleftrightarrow_{ij})^{a^\nleftrightarrow_{ij}}
\end{eqnarray}

\noindent and, by setting $x_i\equiv e^{-\alpha_i}$, $y_i\equiv e^{-\beta_i}$ and $z_i\equiv e^{-\gamma_i}$ \cite{mygrandcanonical},

\begin{eqnarray}
p_{ij}^\rightarrow&\equiv&\frac{x_i y_j}{1+x_i y_j+x_j y_i +z_i z_j},\,p_{ij}^\leftarrow\equiv\frac{x_j y_i}{1+x_i y_j+x_j y_i +z_i z_j}
\label{eq_expreciprinizio}\\
p_{ij}^\leftrightarrow&\equiv&\frac{z_i z_j}{1+x_i y_j+x_j y_i +z_i z_j},\,p_{ij}^\nleftrightarrow\equiv\frac{1}{1+x_i y_j+x_j y_i +z_i z_j}
\label{eq_expreciprfine}
\end{eqnarray} 

\paragraph{The Log-Likelihood Equations.} The log-likelihood function to maximize is 

\begin{eqnarray}
\mathcal{L}(\mathbf{x},\mathbf{y},\mathbf{z})&=&\sum_i\left[ k^\rightarrow_i(\mathbf{A}^*)\ln x_i+k^\leftarrow_i(\mathbf{A}^*)\ln y_i+k^\leftrightarrow_i(\mathbf{A}^*)\ln z_i\right]\nonumber\\
&-&\sum_{i< j}\ln (1+x_i y_j+x_j y_i+z_i z_j)
\end{eqnarray}

\noindent and the values $\mathbf{x}^*$, $\mathbf{y}^*$, $\mathbf{z}^*$ corresponding to the point of maximum can be found by solving the following system

\begin{eqnarray}
\left\{ \begin{array}{ll}
k^\rightarrow_i(\mathbf{A}^*) &= \sum_{j\ne i}\frac{x^*_i y^*_j}{1+x^*_i y^*_j+x^*_j y^*_i+ z_i^* z_j^*}\quad \forall i\\
k^\leftarrow_i(\mathbf{A}^*) &= \sum_{j\ne i}\frac{x^*_j y^*_i}{1+x^*_i y^*_j+x^*_j y^*_i+ z_i^* z_j^*}\quad \forall i\\
k^\leftrightarrow_i(\mathbf{A}^*) &= \sum_{j\ne i}\frac{z^*_i z^*_j}{1+x^*_i y^*_j+x^*_j y^*_i+ z_i^* z_j^*}\quad \forall i
\end{array} \right.
\label{eq_krec}
\end{eqnarray}

\paragraph{Expectation Values and Variances.} The expectation value and variance of $a_{ij}^{\rightarrow}$ (and equivalently for the other dyadic variables) are, respectively, $\langle a_{ij}^\rightarrow\rangle^{*}=(p_{ij}^\rightarrow)^{*}$ and $(\sigma^*[a_{ij}^{\rightarrow}])^2=\langle a_{ij}^{\rightarrow}\rangle^{*}(1-\langle a_{ij}^{\rightarrow}\rangle^*)=(p_{ij}^{\rightarrow})^{*}(1-(p_{ij}^{\rightarrow})^{*})$. Considering that two distinct dyads can be treated as independent random variables and that the second definition of the motifs occurrences only involves products of dyads, their expectation values can be easily computed, as shown in Table \ref{motable2}. The variance of $N_m$ becomes

\begin{equation}
(\sigma^*[N_{m}])^2=\sum_{i,j}\left[\left(\sigma^*[a_{ij}]\frac{\partial N_{m}}{\partial a_{ij}}\right)_{\mathbf{A}=\langle\mathbf{A}\rangle^*}^2+\sigma^*[a_{ij},a_{ji}]\left(\frac{\partial N_{m}}{\partial a_{ij}}\frac{\partial N_{m}}{\partial a_{ji}}\right)_{\mathbf{A}=\langle\mathbf{A}\rangle^*}\right]
\end{equation}

\noindent where now $(\sigma^*[a_{ij}])^2=\langle a^\leftrightarrow_{ij}+a^\rightarrow_{ij}\rangle^*
(1-\langle a^\leftrightarrow_{ij}+a^\rightarrow_{ij}\rangle^*)$ and $\sigma^*[a_{ij},a_{ji}]=\langle a_{ij}^\leftrightarrow\rangle^*-\langle a^\leftrightarrow_{ij}+a^\rightarrow_{ij}\rangle^*
\langle a^\leftrightarrow_{ji}+a^\rightarrow_{ji}\rangle^*$.

\begin{table}[t!]
\centering
\caption{Expectation values of the triadic motifs}
\begin{tabular}{llllll}
\hline\noalign{\smallskip}
Motif $m$ & $\langle N_{m}\rangle_{DCM}$ & $\langle N_{m}\rangle_{RCM}$\\
\noalign{\smallskip}
\hline
\noalign{\smallskip}
$1$ & $\sum_{i\neq j\neq k}(1-p_{ij})p_{ji}p_{jk}(1-p_{kj})(1-p_{ik})(1-p_{ki})$ & $\sum_{i\neq j\neq k}p_{ij}^{\leftarrow}p_{jk}^{\rightarrow}p_{ik}^{\nleftrightarrow}$\\
\hline
$2$ & $\sum_{i\neq j\neq k}p_{ij}(1-p_{ji})p_{jk}(1-p_{kj})(1-p_{ik})(1-p_{ki})$ & $\sum_{i\neq j\neq k}p_{ij}^{\rightarrow}p_{jk}^{\rightarrow}p_{ik}^{\nleftrightarrow}$\\
\hline
$3$ & $\sum_{i\neq j\neq k}p_{ij}p_{ji}p_{jk}(1-p_{kj})(1-p_{ik})(1-p_{ki})$ & $\sum_{i\neq j\neq k}p_{ij}^{\leftrightarrow}p_{jk}^{\rightarrow}p_{ik}^{\nleftrightarrow}$\\
\hline
$4$ & $\sum_{i\neq j\neq k}(1-p_{ij})(1-p_{ji})p_{jk}(1-p_{kj})p_{ik}(1-p_{ki})$ & $\sum_{i\neq j\neq k}p_{ij}^{\nleftrightarrow}p_{jk}^{\rightarrow}p_{ik}^{\rightarrow}$\\
\hline
$5$ & $\sum_{i\neq j\neq k}(1-p_{ij})p_{ji}p_{jk}(1-p_{kj})p_{ik}(1-p_{ki})$ & $\sum_{i\neq j\neq k}p_{ij}^{\leftarrow}p_{jk}^{\rightarrow}p_{ik}^{\rightarrow}$\\
\hline
$6$ & $\sum_{i\neq j\neq k}p_{ij}p_{ji}p_{jk}(1-p_{kj})p_{ik}(1-p_{ki})$ & $\sum_{i\neq j\neq k}p_{ij}^{\leftrightarrow}p_{jk}^{\rightarrow}p_{ik}^{\rightarrow}$\\
\hline
$7$ & $\sum_{i\neq j\neq k}p_{ij}p_{ji}(1-p_{jk})p_{kj}(1-p_{ik})(1-p_{ki})$ & $\sum_{i\neq j\neq k}p_{ij}^{\leftrightarrow}p_{jk}^{\leftarrow}p_{ik}^{\nleftrightarrow}$\\
\hline
$8$ & $\sum_{i\neq j\neq k}p_{ij}p_{ji}p_{jk}p_{kj}(1-p_{ik})(1-p_{ki})$ & $\sum_{i\neq j\neq k}p_{ij}^{\leftrightarrow}p_{jk}^{\leftrightarrow}p_{ik}^{\nleftrightarrow}$\\
\hline
$9$ & $\sum_{i\neq j\neq k}(1-p_{ij})p_{ji}(1-p_{jk})p_{kj}p_{ik}(1-p_{ki})$ & $\sum_{i\neq j\neq k}p_{ij}^{\leftarrow}p_{jk}^{\leftarrow}p_{ik}^{\rightarrow}$\\
\hline
$10$ & $\sum_{i\neq j\neq k}(1-p_{ij})p_{ji}p_{jk}p_{kj}p_{ik}(1-p_{ki})$ & $\sum_{i\neq j\neq k}p_{ij}^{\leftarrow}p_{jk}^{\leftrightarrow}p_{ik}^{\rightarrow}$\\
\hline
$11$ & $\sum_{i\neq j\neq k}p_{ij}(1-p_{ji})p_{jk}p_{kj}p_{ik}(1-p_{ki})$ & $\sum_{i\neq j\neq k}p_{ij}^{\rightarrow}p_{jk}^{\leftrightarrow}p_{ik}^{\rightarrow}$\\
\hline
$12$ & $\sum_{i\neq j\neq k}p_{ij}p_{ji}p_{jk}p_{kj}p_{ik}(1-p_{ki})$ & $\sum_{i\neq j\neq k}p_{ij}^{\leftrightarrow}p_{jk}^{\leftrightarrow}p_{ik}^{\rightarrow}$\\
\hline
$13$ & $\sum_{i\neq j\neq k}p_{ij}p_{ji}p_{jk}p_{kj}p_{ik}p_{ki}$ & $\sum_{i\neq j\neq k}p_{ij}^{\leftrightarrow}p_{jk}^{\leftrightarrow}p_{ik}^{\leftrightarrow}$\\
\hline
\label{motable2}
\end{tabular}
\end{table}

\bibliographystyle{splncs}

\end{document}